\documentclass[twocolumn,showpacs,preprintnumbers,amsmath,amssymb,prl,superscriptaddress]{revtex4}

\usepackage{graphicx}
\usepackage{dcolumn}
\usepackage{bm}

\begin{document}


\title{ $\mu$SR study of the quantum dynamics in the frustrated $S=\frac{3}{2}$ kagom\'e bilayers 
}

\author{D.~Bono}
\author{P.~Mendels}%
\affiliation{%
Laboratoire de Physique des Solides, UMR 8502, Universit\'e Paris-Sud, 91405 Orsay, France
}%

\author{G.~Collin}
\affiliation{
Laboratoire L\'eon Brillouin, CE Saclay, CEA-CNRS, 91191 Gif-sur-Yvette, France
}%

\author{N.~Blanchard}%
\author{F.~Bert}%
\affiliation{%
Laboratoire de Physique des Solides, UMR 8502, Universit\'e Paris-Sud, 91405 Orsay, France
}%

\author{A.~Amato}
\author{C.~Baines}%
\affiliation{%
Paul Scherrer Institut, Laboratory for Muon Spin Spectroscopy, CH-5232 Villigen PSI, Switzerland
}%

\author{A.D.~Hillier}%
\affiliation{%
ISIS Facility, Rutherford Appleton Laboratory, Chilton, Didcot, Oxon, OX11, OQX, United Kingdom
}%

\date{\today}

\begin{abstract}
We present $\mu$SR experiments in the $S=\frac{3}{2}$ kagom\'e bilayer compound Ba$_{2}$Sn$_{2}$ZnGa$_{10-7p}$Cr$_{7p}$O$_{22}$ (BSZCGO$(p)$) and compare it to the isostructural SrCr$_{9p}$Ga$_{12-9p}$O$_{19}$ (SCGO$(p)$), including for the latter new results for $p\geq0.89$. Quantum-dynamical low energy magnetic excitations are evidenced in this novel compound. We study the evolution of the muon relaxation rate with $p$, $T$ and field. A phenomenological model for the muon relaxation based on sporadic dynamics due to spin excitations in a singlet sea proposed by Uemura \emph{et al.} [Phys. Rev. Lett. \textbf{73}, 3306 (1994)]. is extended to all fields and $T$-range. Its connexion to the RVB picture is discussed, and we argue that such coherent states might mediate the interactions between ``impurities'' which induce the spin glass freezing.
\end{abstract}


\maketitle

The stabilization of a spin liquid state, in which all kinds of magnetic correlation functions are short ranged, seems now granted theoretically in the $S=\frac{1}{2}$ Heisenberg kagom\'e lattice with nearest neighbour (NN) antiferromagnetic interactions. The ground state (GS) is probably of resonating valence bond type (RVB), built from a macroscopic number of singlet states \cite{Mambrini00}. If any, a gap between the GS and the first triplet state is expected to be fairly small ($\sim J/20$) \cite{Waldtmann98} in comparison with the exchange constant $J$. Most striking, due to both corner sharing geometry and half integer spins, an exponential density of low lying singlet excitations, with energies smaller than the gap, is predicted \cite{Mila98} and yields an extensive entropy at $T=0$ \cite{Misguich03}. Finally, the nature of the first magnetic excitations, such as unconfined spinons, are still under debate \cite{Misguich03}. 

Among all corner sharing highly frustrated magnets, only a few are good candidates fulfilling the important conditions of the pure Heisenberg lattice with NN couplings. The combination of the weakness of the single-ion anisotropy and of a direct overlap exchange are certainly the major advantages of the recently discovered chromate $S=\frac{3}{2}$ kagom\'e bilayer BSZCGO$(p)$ \cite{Hagemann01,BonoRMN} and the long studied SCGO$(p)$ \cite{Limot02}, where $p$ is the Cr$^{3+}$ network covering rate. Beyond the absence of ordering well below the Curie-Weiss temperature, the unusual large value of the specific heat unveils a high density of low lying excitations \cite{Ramirez90,Hagemann01} and its field independence suggests that the excited states are mostly singlets \cite{Ramirez00}. Moreover, their GS is found essentially fluctuating as proven by neutron experiments \cite{Broholm90}, and $\mu$SR for SCGO$(p)$ \cite{Uemura94,Keren00}, although an intrinsic spin glass (SG) signature is observed in susceptibility measurements \cite{Schiffer97,Limot02}. The origin of such a SG state in a disorder-free system still awaits for a complete understanding although recent theoretical approaches catch some of the facets of this original GS~\cite{Mila02}.

In this context BSZCGO$(p)$ appears as a unique compound~: it has a more pronounced 2D character than SCGO$(p)$, with a larger interbilayer distance \cite{Hagemann01} and \emph{all} the magnetic sites lying on a corner sharing lattice~\footnote{On the contrary, SCGO$(p)$ has Cr pairs in between the bilayers.}. In addition, a twice lower SG temperature $T_{g}$ has been reported \cite{Hagemann01} and a new type of defects other than non-magnetic substitutions (likely bond defects), dominates the low-$T$ susceptibility \cite{BonoRMN}. This offers the possibility of probing the influence of $T_{g}$ and the relevance of dilution on the evolution of the $T\rightarrow0$ dynamics, by a simple comparison of two quasi-identical kagom\'e bilayers.

In this Letter, we present the first Muon Spin Relaxation ($\mu$SR) study of the spin dynamics in BSZCGO$(p)$ and revisit the case of SCGO$(p)$ for low dilutions now available. This technique has proven to be a front tool for the study of quantum dynamical states in a vast family of singlet GS systems. Whereas specific heat is sensitive to all kinds of excitations, including likely dominant low lying singlets at low-$T$ \cite{Ramirez00}, muons ($\mu^{+}$, $S=\frac{1}{2}$) probe \emph{magnetic} excitations \emph{only}. Their relaxation rate $\lambda$ is linked to the spin-spin time correlation function $\lambda=1/T_{1}\sim \int_{0}^{\infty}\langle \mathbf{S}(0)\cdot\mathbf{S}(t)\rangle \cos(\gamma_{\mu} H_{LF} t) dt$, where $H_{LF}$ is the external field applied along the $\mu^{+}$ initial polarization and $\gamma_{\mu}$ is the $\mu^{+}$ gyromagnetic ratio. The typical time window (10~ns-10~$\mu$s) 
allows to observe spin fluctuations which set in the specific range of frequencies ($\sim10^{9}$~Hz) in between NMR and neutron experiments. Not only in SCGO \cite{Uemura94,Keren00} and in other kagom\'e systems \cite{Keren96,Fukaya03} but also in other spin-singlet compounds \cite{Kojima95}, the signature of the quantum dynamics when $T\rightarrow0$ could be identified to the occurence of a plateau of $\lambda$ and a field-independent initial Gaussian shape of the $\mu^{+}$ polarisation $P_{z}(t)$. Extending the original guess of Uemura \emph{et al.} to explain this exotic fluctuating GS 
\cite{Uemura94}, we propose for the first time a phenomenological model which explains both $T$- and field-dependence of $P_{z}(t)$ for all $p$ in these kagom\'e bilayers.

      \begin{figure}[tbp!] \center
\includegraphics[width=1\linewidth]{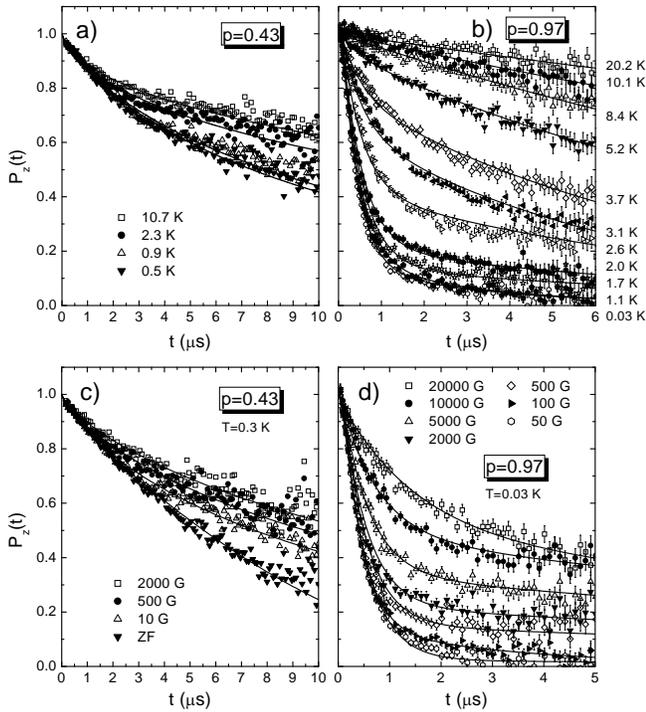}
      \caption{ \label{fig_Asym_T} a,b)~$T$-dependence of the $\mu^{+}$ polarization $P_{z}(t)$ in BSZCGO under weak (10~G for $p=0.43$ and 100~G for $p=0.97$) longitudinal field. Notice the different time scales for both samples. c,d)~$H_{LF}$-dependence at base-$T$. The line for $p=0.43$, ZF, is a square-root exponential multiplied by the Kubo-Toyabe function in zero field ($\Delta_{ND}\sim0.08\mu$s$^{-1}$), to take into acount the nuclear dipole contribution. The other lines are fits described in the text.}
      \end{figure}

In order to illustrate the evolution of the properties with dilution, we first focus on the qualitative differences (Fig.~\ref{fig_Asym_T}) between two typical low and high occupations $p$ of the kagom\'e bilayers. Starting from high-$T$ ($T>6$~K in Fig.~\ref{fig_Asym_T}(a,b)), a conventional paramagnetic behavior is found for both cases, with an exponential variation of $P_{z}(t)$ in the dense case and square root exponential in the dilute case. The evolution of $P_{z}(t)$ at low-$T$ is markedly different. For $p=0.97$, the relaxation rate increases by more than two orders of magnitude to reach a $T$-independent value for $T\lesssim T_{g}\approx1.5$~K \cite{Hagemann01} (Fig.~\ref{fig_Asym_T}(b)). For $T=0.03$~K, $P_{z}(t)$ displays a shape in between exponential and Gaussian at early times and reaches zero value at long times (Fig.~\ref{fig_Asym_T}(d)). It also displays a much weaker field dependence than for a frozen magnetic state. Indeed, for $\lambda\sim2~\mu$s$^{-1}$, a 500~G field would completely decouple the $\mu^{+}$ polarisation \cite{Uemura94}. Following the arguments initially developed for SCGO$(p)$ \cite{Uemura94,Keren00}, all these properties are the first indications of a quantum dynamical magnetic state in BSZCGO$(0.97)$. For $p=0.43$, we also find a dynamical state but, at variance with the previous case, only a weak $T$-dependence is observed. Also, the polarization displays a square root exponential decay for all $T$ (Fig.~\ref{fig_Asym_T}(a)) and for any field $H_{LF}$ (Fig.~\ref{fig_Asym_T}(c)). A weak plateau of the relaxation rate is still observed below $0.5$~K. However the relaxation rate increases weakly but continuously when $T$ decreases for $p=0.3$, with a very weak $H_{LF}$-dependence. This is typical of the pure paramagnetic fast fluctuations limit for diluted magnetic systems, although strongly frustrated antiferromagnetic interactions are still reported \cite{BonoRMN}. 

      \begin{figure}[tbp!] \center
\includegraphics[width=0.9\linewidth]{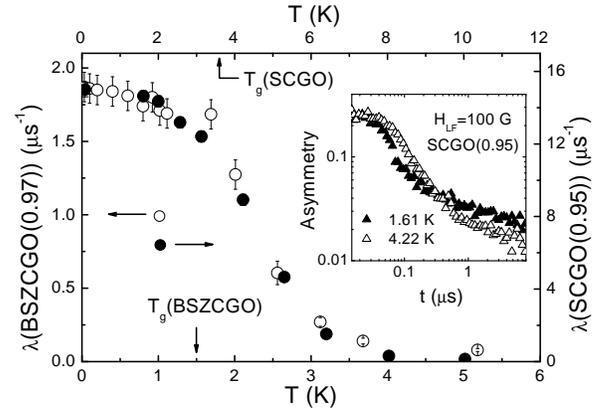}
      \caption{ \label{fig_lambda_scale}  Comparison of the evolution of the $\mu^{+}$ relaxation rate $\lambda$ \emph{vs.} $T$ in BSZCGO(0.97) and SCGO(0.95), from a $1/e$ analysis. Inset~: recovery of a part of the asymmetry at low-$T$ and long times in SCGO below $T_{g}$ (log-log scale).
       }
      \end{figure}
      
In a first step of analysis, we estimate the $\mu^{+}$ relaxation rate $\lambda$ using the $\frac{1}{e}$ point of $P_{z}(t)$, as discussed in \cite{Keren00}. This allows to single out information about two important issues using a simple model-independent analysis~: the role of the SG-like transition and the impact of the non-magnetic dilutions on the dynamics.

From high-$T$, $\lambda$ increases by more than two orders of magnitude down to $T_{g}$ for both purest samples, to reach the relaxation rate plateau $\lambda_{T\rightarrow0}$ for $T\lesssim T_{g}$. With a $T$-scale twice larger for SCGO(0.95) than in BSZCGO(0.97), $\lambda (T)$ perfectly scales on the $T$ axis as shown in Fig.~\ref{fig_lambda_scale} \footnote{We attribute the factor 8 between $\lambda$ in both systems to a different coupling of the $\mu^{+}$ to the Cr spins. Transverse field experiments show for instance a larger linewidth in SCGO(0.95) than in BSZCGO(0.91) (D.~Bono \emph{et al.}, \emph{unpublished}) despite a lower defect term \cite{BonoRMN}.}. This ratio is very close to their $T_{g}$ ratio of 2.3(2), which points to a link between the formation of the SG-like state and the presence of quantum fluctuations. This seems to be a quite common feature of various systems with singlet GS systems \cite{Uemura94,Kojima95,Keren96,Keren00,Fukaya03}.

\begin{figure}[tbp!] \center
\includegraphics[width=0.9\linewidth]{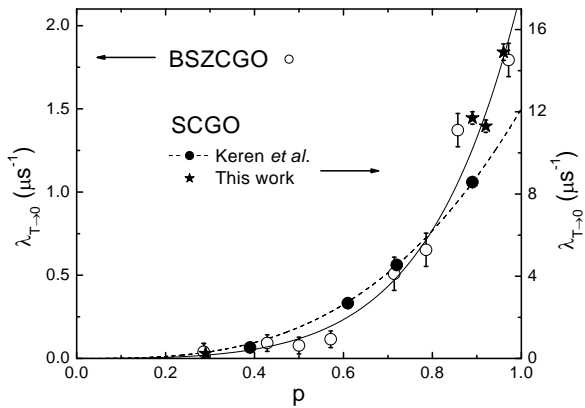}
      \caption{ \label{fig_lambda_p}  $p$-dependence of $\lambda_{T\rightarrow0}$ for BSZCGO (open symbols) and SCGO (full symbols). Continuous (dashed) lines are $\sim p^{3}+1.9(3)p^{6}$ ($\propto p^{3}$, \cite{Keren00}) fits.
       }
      \end{figure}

In Fig.~\ref{fig_lambda_p} we report the variation of $\lambda_{T\rightarrow0}(T)$ for various BSZCGO$(p)$ samples and compare them to SCGO$(p)$ \cite{Uemura94,Keren00}. Although additionnal $p$-independent defects are dominant in BSZCGO$(p)$ and evidenced from a large Curie-like term in macroscopic susceptibility at low-$T$ \cite{BonoRMN}, it is very surprising to find a very similar quantitative $\lambda_{T\rightarrow0}$ variation in the \emph{whole} $p$ range for both systems. Therefore we conclude that \emph{only} defects related to \emph{dilution} influence the relaxation rate.

In a classical framework, a coplanar arrangement with zero-energy excitation modes, involving spins on hexagons, was proposed in the literature to be selected at low-$T$ \cite{Chubukov92}. Inelastic neutron scattering experiments on the spinel compound ZnCr$_{2}$O$_{4}$ are the first recent indication in favor of such excitations \cite{Lee02}. Therefore, one expects the $\mu^{+}$ relaxation to scale with the number of fully occupied hexagons, $\propto p^{6}$. The BSZCGO$(p)$ data altogether with our SCGO$0.89\leq p\leq0.95$ samples indicate that $\lambda_{T\rightarrow0}(p)$ is well accounted for by adding a dominant $p^{6}$ term to the $\propto p^{3}$ term proposed for SCGO($p\leq0.89$) in \cite{Keren00} (Fig.~\ref{fig_lambda_p}). This certainly indicates that the relaxation is not induced by single spin excitations but rather \emph{collective} excitation processes extending at least on triangles and hexagons.

One of the pending challenges in the analysis of $\mu$SR in kagom\'e frustrated antiferromagnets is to find a theoretical model reproducing $P_{z}(t)$ for \emph{all} fields and \emph{all} temperatures. In the quantum framework of a singlet GS, one needs unpaired spins in order to generate magnetic excitations responsible for the $\mu^{+}$ relaxation. Such excitations can be ascribed by spin $\frac{1}{2}$ unconfined spinons. Their location vary in time with no loss of coherence of the excited state. Hence, a given $\mu^{+}$ couples to a spin only a short fraction $ft$ of the time $t$ after implantation and one can use a model of ``sporadic'' field fluctuations to describe $P_{z}(t)$. This was the initial guess by Uemura \emph{et al.} to account for the weakness of the field dependence of $P_{z}(t)$ at low-$T$ in SCGO(0.89) \cite{Uemura94}. This polarization is given by a ``sporadic'' dynamical Kubo-Toyabe function which can be approximated by the Keren analytical function $P_{z}^{K}(ft,\Delta,H_{LF},\nu)=P_{z}^{K}(t,f\Delta,fH_{LF},f\nu)$, valid in the limit $\nu\geq\Delta$ \cite{Keren94}, quite appropriate here. $\Delta/\gamma_{\mu}$ is the local field created on the $\mu^{+}$ site by a paramagnetic neighbouring released spin, with a corresponding exponential time correlation function, $\exp(-\nu t)$, characterized by a fluctuation frequency $\nu$. $\Delta$ is related to the $\mu^{+}$ location in the unit cell and its value is therefore $p$-independent for a given system but is expected to vary from BSZCGO to SCGO. The Gaussian at early times, the weakness of the field dependence and a dynamical relaxation down to 0 are altogether related to the $f$ factor and $\nu\sim \Delta$. In addition to this sporadic relaxation, we found that a more conventional Markovian relaxation channel needs to be introduced in order to fit the long times tail ($t\gtrsim3~\mu$s) for all fields and $T$, which yields an exponential variation of $P_{z}(t)$. We therefore write
\begin{equation}
\label{eqmodel}
	P_{z}(t)=xP_{z}^{K}(t,f\Delta,fH_{LF},f\nu)+(1-x)e^{-\lambda^{\prime} t} \ ,
\end{equation}
where $x$ represents the weight of the short time sporadic dynamical function, associated with the spinons dynamics. In the following, we first detail how all these parameters can be reliably deduced from the data and sketch a physical picture consistent with our results. 

      \begin{figure}[tbp!] \center
\includegraphics[width=0.9\linewidth]{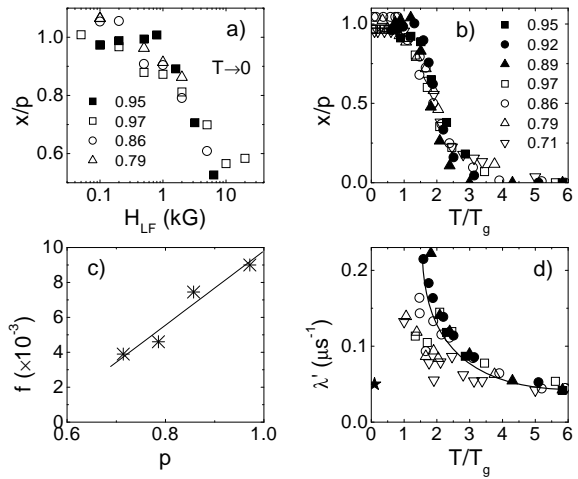}
      \caption{ \label{fig_fitx} Fitting parameters of Eq.~\ref{eqmodel}. a,b)~$H_{LF}$- and $T$-dependence of $x$ for BSZCGO$(p)$ (open symbols) and SCGO$(p)$ (full symbols). c)~$p$-dependence of $f$. d)~$T$-dependence of $\lambda^{\prime}$. 
       }
      \end{figure}

We first present our analysis for $T\ll T_{g}$ in BSZ\-CGO$(p)$. In order to limit the number of free parameters, we make the minimal assumption that the external field does not influence the dynamics of the coherent spinon term. $\nu$ and $\Delta$ values are shared for all $p$ and $H_{LF}$ and are set by our data taken at low fields ($H_{LF}<500$~G) for various $p$. $f$ is adjusted for each $p$ and its variation accounts for the evolution of $\lambda_{T\rightarrow0}(p)$. For low fields, $x$ is found close to 1 for the purest samples (Fig.~\ref{fig_Asym_T}(a,b)), making $\lambda^{\prime}$ a non-relevant fitting parameter. On the contrary, the high field data enable us to track the evolution of $x$ with $H_{LF}$ and to determine a value of $\lambda^{\prime}$. We could not extend the fits below $p=0.71$ since the weak field dependence prevents an unambiguous determination of the parameters.

We find perfect fits of our data (Fig.~\ref{fig_Asym_T}(d)) with $\nu\sim 1000~\mu$s$^{-1}$, $\Delta\sim350~\mu$s$^{-1}\sim\gamma_{\mu}\times4$~kG and an average value of $f\sim0.006$. We find a nearly flat $p$-independent long time relaxation rate for $T\ll T_{g}$ ($\lambda^{\prime}\sim0.05~\mu$s$^{-1}$, $\bigstar$ in Fig.~\ref{fig_fitx}(d)). An important finding is that $x$ is of the order of $p$ at low fields (Fig.~\ref{fig_fitx}(a,b)). This may be related to theoretical computations showing that the correlations are enhanced around spin vacancies in the kagom\'e lattice \cite{Dommange03}, which would destroy locally the liquid state. Besides, $x$ decreases appreciably for $H_{LF}\sim10$~kG whatever the value of $p$. We can associate this decrease to the existence of an energy scale $\sim$ 1~K, which is of the order of $T_{g}$. Finally, we observe a linear variation of $f$ with $p$ (Fig.~\ref{fig_fitx}(c)) and $f$ tends to vanish around $p\sim0.5$, a limit consistent with our classical approach (Fig.~\ref{fig_lambda_p}). This indicates that even far from the substituted sites, the coherent state is somehow affected, e.g the density of spinons could be smaller.

As suggested by the similar $p$-variation of $\lambda_{T\rightarrow0}$ in both BSZCGO and SCGO, we assume that the excitation modes are identical in both systems, i.e., $f$ and $\nu$ are kept the same for comparable $p$. We find $\Delta\sim1200~\mu$s$^{-1}$ for our SCGO samples ($p\geq0.89$), in agreement with previous work on SCGO(0.89) \cite{Uemura94}. 
In order to cross-check these values of $\Delta$, we use the high-$T$ ($\sim50$~K) data for BSZCGO (SCGO), $\lambda\sim0.01~\mu$s$^{-1}$ ($0.03~\mu$s$^{-1}$). In the appropriate paramagnetic limit, the fluctuation rate can be estimated from $\nu\sim\sqrt{z}JS/k_{B}\hbar\sim2\times10^{13}$~s$^{-1}$ \cite{Uemura94} using a coupling $J=40$~K \cite{BonoRMN} and an average number $z=5.14$ of Cr NN \cite{Limot02}. We extract $\Delta=\sqrt{\lambda\nu/2}\sim300~\mu$s$^{-1}$ ($600~\mu$s$^{-1}$), consistent with our fits. It is quite rewarding to find that a \emph{common} physical picture underlies all the sets of data at low-$T$ for \emph{both} kagom\'e bilayers.   

We now extend the fits of $P_{z}(t)$ to the whole $T$-range (Fig.~\ref{fig_Asym_T}(b)), fixing $f$ to its low-$T$ value in order to limit the number of free parameters. As expected from the change of shape around $T_{g}$ \cite{Uemura94,Keren00}, the weight $x$ of the sporadic term decreases to finally enter a high-$T$ regime (Fig.~\ref{fig_fitx}(b)) where the $\mu^{+}$ relaxation becomes exponential ($x\rightarrow0$). Notice the similarity between the field and $T$-dependence of $x$ which indicates that the sporadic regime is destroyed when the energy is of the order of $T_{g}$. Fig.~\ref{fig_lambda_p}(b) displays a sharp crossover from one state to the other, between $T_{g}$ and $ \sim 3T_{g}$, corresponding to the $\lambda$ steep decrease (Fig.~\ref{fig_lambda_scale}), for both BSZCGO$(p)$ and SCGO$(p)$. For $T>T_{g}$, $\nu$ and $\lambda^{\prime}$ (Fig.~\ref{fig_fitx}(d)) decrease by one order of magnitude up to $\sim10T_{g}$. 

At high-$T$, it is natural to think in terms of paramagnetic fluctuating spins. Lowering $T$, $\lambda^{\prime}$ seems to diverge at $T_{g}$ and below, the weight of the exponential term at low fields ($1-x\sim 1-p$) could correspond to localized frozen spins, reflecting a glassy component. It is noteworthy that $T_{g}$ does not increase with $1-p$ but rather decreases, at the opposite of the case of canonical spin glasses. We could further confirm the existence of such a frozen component in SCGO$(p=0.95-0.89)$, since a clear recovery of a small part ($\sim 4$\%) of the asymmetry is found at long times (e.g. Fig.~\ref{fig_lambda_scale}(inset)). It is observed for $t\sim100\,\lambda^{-1}$, which is out of the experimental range in the case of BSZCGO$(p)$.

To summarize, the picture based on a coherent RVB state, which magnetic excitations are mobile fluctuating spins $\frac{1}{2}$ on the kagom\'e lattice, explains well the data, provided that (i)~these excitations can be generated even for $T\rightarrow0$. This underlines the smallness of the ``magnetic'' gap, if any, typically $\Delta < J/1000$~; (ii)~an energy scale related to $T_{g}$ (fields of the order of 10~kG or, equivalently, temperatures of the order of $T_{g}$, which vary very little with $p$) are high enough to destroy the coherent RVB type state~; (iii)~the substitution defects are accounted for by an additional classical relaxation process.

The absence of energy scale for spin $\frac{1}{2}$ excitations could explain quite well why the transition to the SG state is fairly independent of $p$. Indeed, in this framework, spinons could mediate the interactions between magnetic defects localized around spin vacancies. This is corroborated by all the scaling properties in $T/T_{g}$. The transition to the SG state corresponds then to the formation of the coherent singlet state rather than any interaction strength between defects. 

We thank A.~Keren, C.~Lhuillier, G.~Misguich and R.~Moessner for fruitful discussions. This work was performed at the S$\mu$S, PSI, Villigen CH, and at the ISIS facility, RAL, Didcot UK.

\end{document}